\documentclass[twocolumn]{revtex4-1}

\usepackage[utf8]{inputenc}
\usepackage{amsmath,amssymb,mathtools}
\usepackage{lmodern,bbm,graphicx,float}
\usepackage{comment,textcomp}
\usepackage{hyperref,url,appendix}

\begin{document}

\title{Stable CW Operation of Nb3Sn SRF Cavity at 10 MV/m using Conduction Cooling}
\author{N. Stilin}
\email{nas97@cornell.edu}
\author{A. Holic}
\author{M. Liepe}
\author{R. Porter}
\author{J. Sears}
\date{\today}

\begin{abstract}
    Superconducting radio-frequency (SRF) accelerating cavities are a promising technology for compact, high-power, MeV-scale accelerators, providing  continuous beam operation for a wide range of potential applications in industry, medicine, national security, and science. However, SRF cavities have traditionally been cooled using complex and expensive cryogenic infrastructure, which has been a significant obstacle in employing this powerful technology  in small-scale systems. With the recent progress on high-performance  Nb$_3$Sn-coated SRF cavities and the introduction of cryocoolers, which can provide turn-key style conduction cooling below 4.2~K, this barrier can now be overcome. At Cornell University, we have developed a  prototype setup that utilizes a commercial cryocooler to provide the necessary cooling for operation of a 2.6~GHz Nb$_3$Sn-coated SRF cavity. We have now demonstrated first stable  operation in continuous mode at accelerating fields up to 10~MV/m with a quality factor of Q$_0$~=~4x10$^9$, thereby achieving an important milestone in making turn-key SRF a reality.
\end{abstract}
\maketitle

\section{Introduction}

Previous studies have shown that there is a wide range of potential industrial, medical, national security, and science applications for small-scale accelerators operating in continuous wave (CW) mode with energy ranges of a few MeV. For example, direct or indirect radiation (via production of X-rays) by electron beams at such energies could be used for wastewater treatments, medical  sterilization, treatment of asphalt for increased durability, and so on \cite{argonne,DOE_AAF}. While SRF cavities are an attractive solution for such accelerators thanks to their high fields and high efficiency in CW operation with beam currents up to the Ampere-scale, one challenge this technology faces is the need to be cooled to cryogenic temperatures of a few Kelvin. Up until the last few years, niobium cavities have been the focus of SRF research and technology. With a superconducting transition temperature ($T_c$) of about 9.2~K, niobium cavities frequently must be cooled to temperatures of around 2~K to achieve the efficiency sought for most applications \cite{hasan}. Such cooling  requires substantial cryogenic infrastructure, involving complex and expensive  systems that use liquid helium and nitrogen. For small-scale operations, the need for this scale of infrastructure can prove to be a daunting obstacle. Not only do they face up-front costs of constructing cryogenic systems, but also ongoing costs and required expertise for continuous operation and maintenance.

This brings our attention to the use of Nb$_3$Sn-coated SRF cavities. Nb$_3$Sn has a $T_c$ just over 18~K, which means it can operate efficiently at temperatures of 4~K or even higher \cite{PhysRevSTAB.17.112001}. While still quite low, this temperature range has the advantage of enabling an alternative cooling scheme based on conduction cooling from cryocoolers.  Recent advancements in cryocooler technology has made systems available which can dissipate up to 2~W of heat at temperatures of 4.2~K. These cryocooler systems can be considered ``dry cryostats,'' as they require no nitrogen or liquid helium to operate. While their cooling efficiency is low compared to that of large-scale liquid helium cryogenic cooling plants, their significantly lower operational costs and simple turn-key style control makes them much more accessible, as is critical for small-scale accelerator applications. First demonstrated by research conducted at Cornell University in recent years, Nb$_3$Sn cavities are now able to operate with accelerating fields well above 15~MV/m with low heat dissipation at 4.2~K  \cite{PhysRevSTAB.17.112001,nb3sn}. This indicates that Nb$_3$Sn cavities could be used to operate efficiently with accelerating fields matching the requirements of the small-scale applications mentioned previously, while also being cooled by cryocoolers rather than specialized cryogenic liquid helium infrastructure. If developed successfully, this type of operation would help pave the way to making SRF technology accessible to a much wider range of applications than it is today.

At Cornell, we have implemented the use of a cryocooler in a new conduction cooling assembly and have started tests on a 2.6~GHz Nb$_3$Sn cavity. We believe that using a higher-frequency cavity like this one provides an additional advantage of compactness, as overall instrument size could be an important consideration in certain future applications. Before moving on, we would like to acknowledge the recent reports by Fermilab \cite{dhuley2020demonstration} and Jefferson Lab \cite{jlab}, in which stable operation using cryocoolers with Nb$_3$Sn cavities at lower frequencies and lower fields were demonstrated.

\section{Apparatus}

In this setup we used a Cryomech PT420-RM cryocooler, which is rated for cooling 1.8~W at 4.2~K. This is slightly lower than the 2.0~W rating of the standard PT420 model, but has the advantage of its motor operating remotely from the rest of the cooling assembly. This was chosen with the intent of reducing  vibrations coupled to the main cavity assembly. 

\begin{figure}[h]
	\floatstyle{boxed}
	\begin{center}
		\includegraphics[width=1\linewidth]{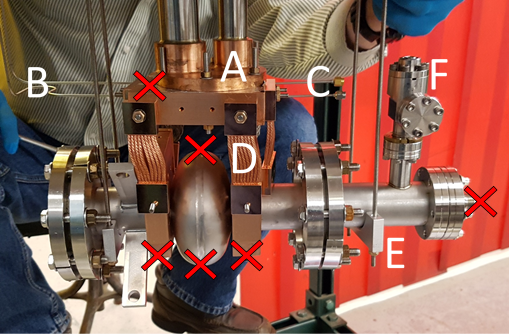}
		\break
		\includegraphics[width=1\linewidth]{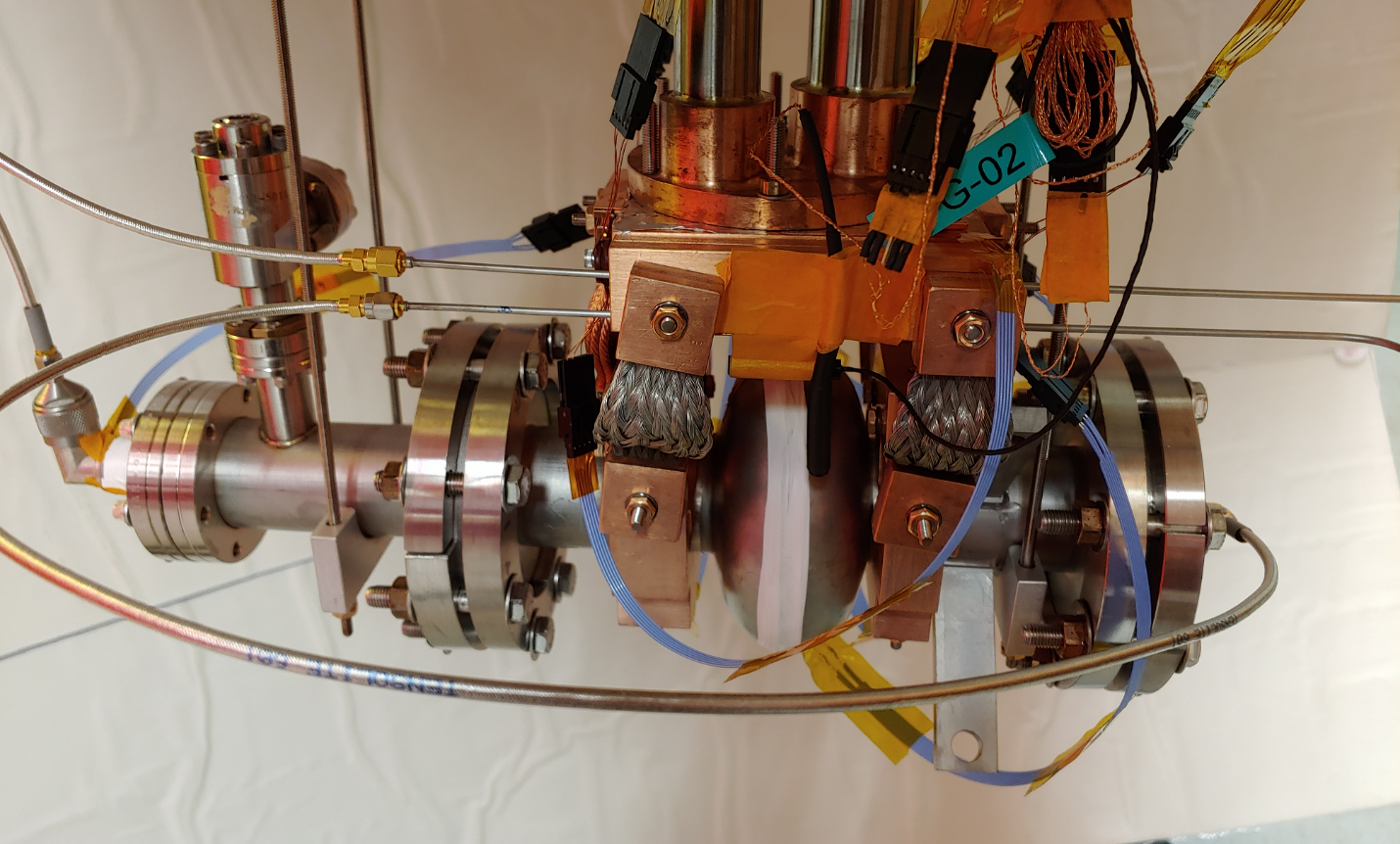}
	\caption{ABOVE: 2.6~GHz Nb$_3$Sn cavity assembly with  conductive cooling anchor  to the second stage of the cryocooler. Red X marks indicate locations of six Cernox sensors:  (clockwise from upper-left) 2nd stage, equator top, RF input coupler, thermal beam  pipe clamp at input coupler side, equator bottom, second thermal beam pipe clamp. Other labelled elements are: (A) 2nd stage cold plate, (B) semi-rigid RF cables, (C) connectors that transition to semi-flexible cables, (D) braided copper thermal straps which connect beam tube clamps to second stage, (E) aluminum cradles used to bear the weight of the assembly, and (F) all-metal valve used for pumping down the cavity. BELOW: Cernox sensors, magnetic flux gates, and semi-flexible cables added to the assembly.}
	\label{cavity}
	\end{center}
\end{figure}

The primary cavity assembly was thermally connected to the second stage cold head for heat dissipation during operation. Our setup used copper clamps made in-house which were secured to the cavity beam tube right next to the irises. To provide better thermal contact, indium was rolled into thin strips and wrapped around the beam tubes before attaching the clamps. The clamps were then anchored to the second stage using thermal straps made of braided copper. To eliminate excessive strain on the copper straps, small aluminum cradles hung from a support ring at the first stage were used to bear the weight of the cavity assembly. The primary cavity assembly, including some of the instrumentation attached, can be seen in Fig \ref{cavity}. The entire cavity assembly was then wrapped in 10 layers of aluminized mylar superinsulation.

The cavity assembly was surrounded by cylindrical cans made of thin copper and mumetal sheets, which were thermally anchored to the first stage cold head of the cryocooler at the top; these acted as thermal and magnetic shields. In addition, the first stage cold head was used to thermally anchor both the RF power cables (semi-rigid cables from Carlisle Interconnect) and various sensor cables. The copper can and cable connections were covered in 30 layers of aluminized mylar.

A total of eight Cernox sensors and two magnetic flux gates were used during our tests. Two Cernox sensors were used to monitor the 40~K stage, with one directly on the first stage and one attached to the bottom of the copper thermal shield. The remaining six were located at various points on the cavity assembly: second stage, forward power coupler, beam tube clamps, and cavity equator; see  Fig \ref{cavity}. The cables from these six Cernox temperature sensor and the RF power cables were again thermally anchored to the second cryocooler stage before connecting to the cavity assembly.   Two magnetic flux gates were placed near the cavity equator, oriented orthogonal to each other to monitor magnetic fields near the cavity during cool down. 

\section{Initial Results}

\begin{figure}[h]
	\floatstyle{boxed}
	\begin{center}
		\includegraphics[width=1\linewidth]{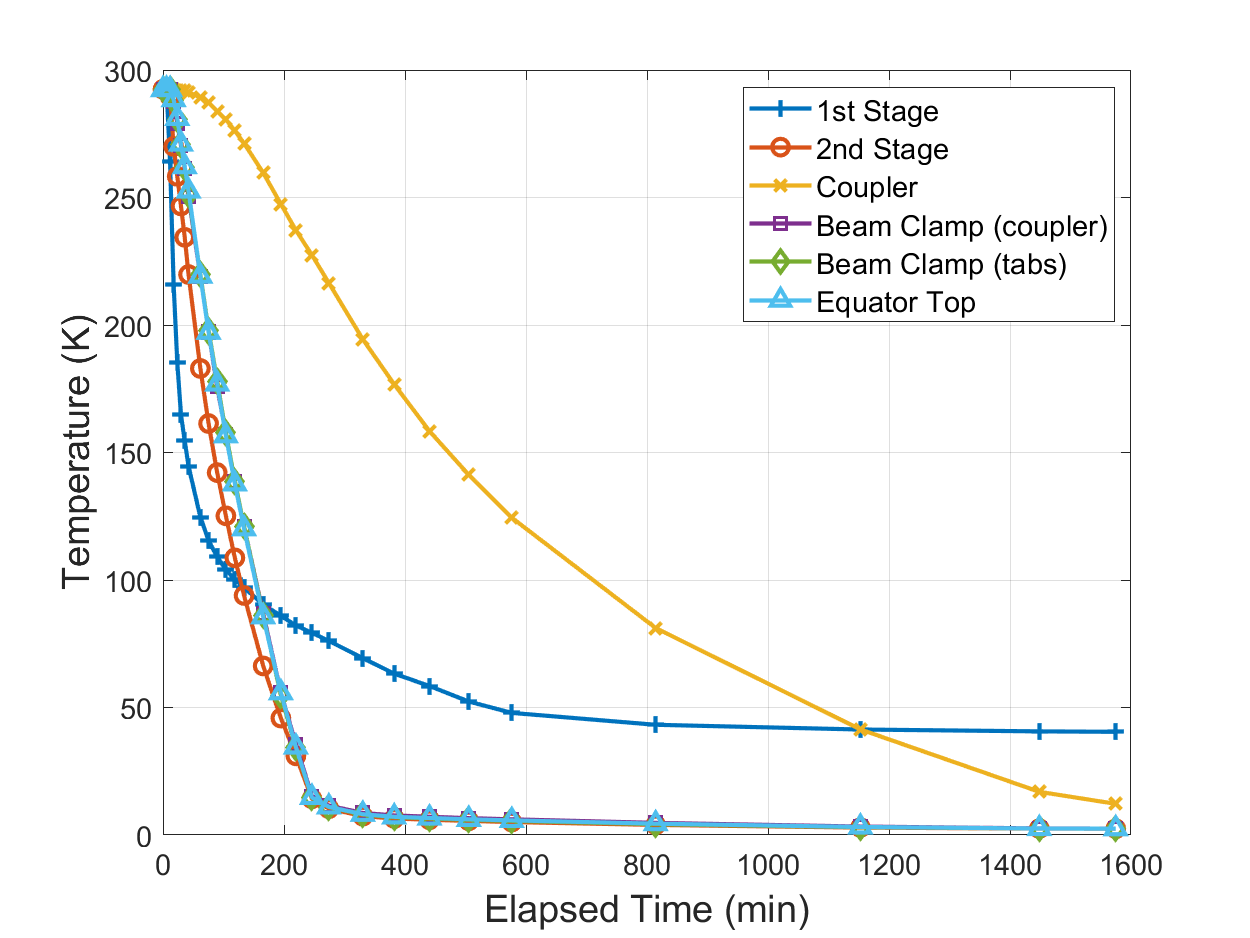}
		\caption{Initial cooldown of cavity assembly from room temperature. Data from the bottom copper thermal shield  and bottom cavity equator are excluded due to the sensors malfunctioning. At this scale, the cavity and beam clamp temperatures are nearly identical and appear completely overlapped on the graph. Once cooled to around 10~K, the second stage temperature also overlaps.}
		\label{cool}
	\end{center}
\end{figure}

The first set of data presented in this report was taken after the assembly initially cooled down from room temperature. A plot of the temperatures during the cooldown is shown in Fig. \ref{cool}. In this first test, the cavity reached an accelerating field of just under 8~MV/m, with the cavity dissipating about 2~W. At low fields the quality factor was 2x10$^9$, which fell below 10$^9$ at higher fields. The ambient magnetic fields near the cavity were measured to be just under 10~mG. During this first test, measurements were made quickly after turning on RF power at high fields, such that we were not performing ``continuous'' operation. This was done due to a concern about heating by the power dissipated in the semi-rigid and semi-flexible cables at higher fields. Beyond 5~MV/m the input coupling factor began dropping noticeably, and at 8~MV/m over 10~W of power was put into the cables.

\begin{figure}[h]
	\floatstyle{boxed}
	\begin{center}
		\includegraphics[width=1\linewidth]{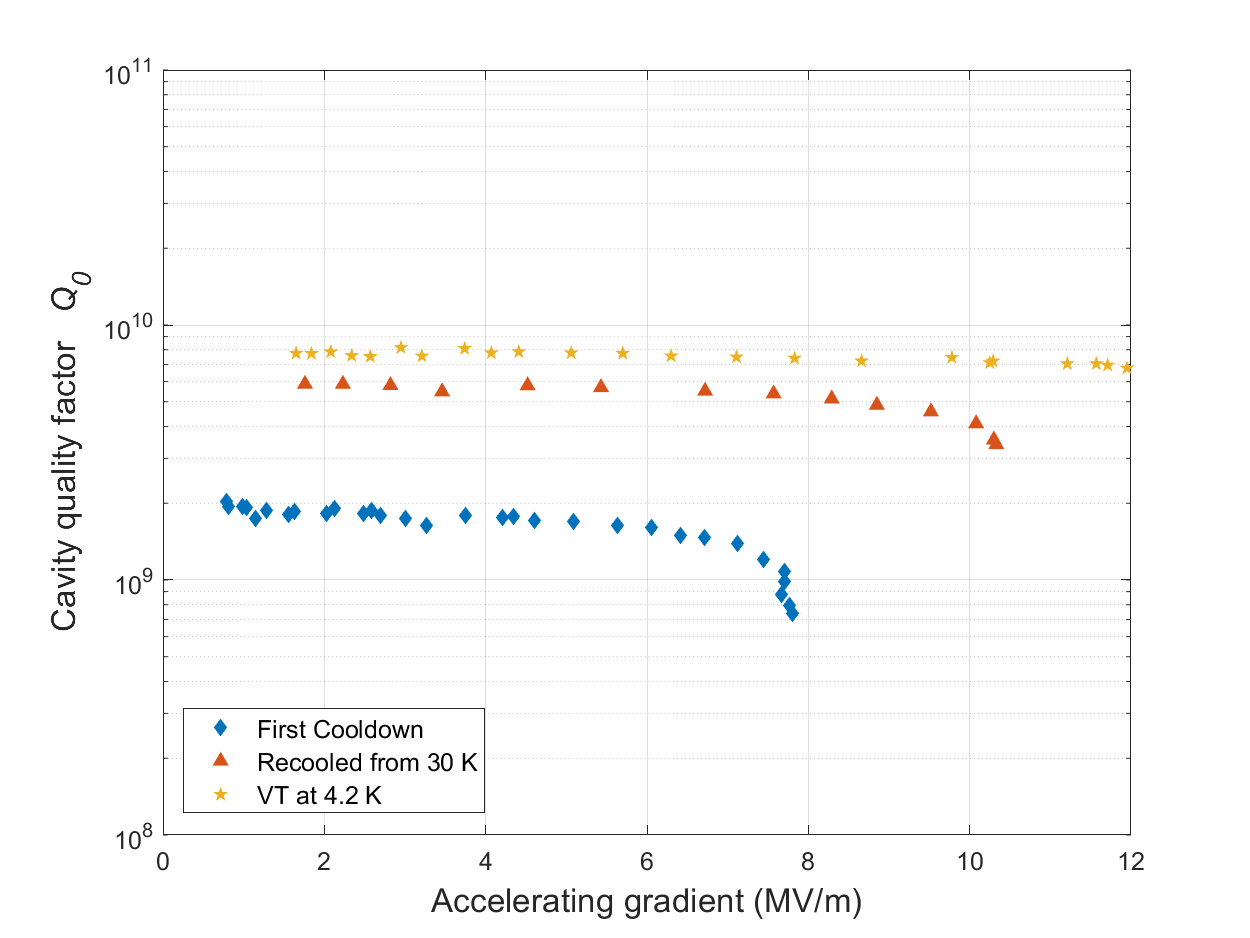}
		\caption{Comparison of quality factor vs accelerating gradient for a 2.6~GHz Nb$_3$Sn cavity from three different tests. First set is from a cavity test with cryocooler based conduction cooling done immediately after the initial cooldown from room temperature. Second set is from a cryocooler based cavity test done after temperature cycling to 30~K and back. Third set is from a baseline vertical test (VT) of the same cavity in a liquid helium bath at 4.2~K.}
		\label{Qcompare}
	\end{center}
\end{figure}

After the initial test a couple rounds of temperature cycling were performed, in which the cavity was allowed to warm up above $T_c$ before cooling down again. This was done to examine how thermal gradients across the cavity might be improved during cooldown through T$_c$, which could in turn improve RF performance. Temperature cycling to 30~K produced the best results, with the cavity reaching an accelerating field of 10.3~MV/m after the thermal cycle and dissipating just under 1~W. The input coupling factor was much better in this test, such that only 1.4~W was put into the cables at the highest field reached. This allowed for properties in continuous operation to be measured, in which the cavity temperatures were allowed to stabilize before taking measurements. At higher fields, this meant that the RF power was kept on for several minutes at a time. In this test, the  measured  quality factor reached 6x10$^9$ at low fields and 4x10$^9$ at 10~MV/m. The results from the two tests described, along with data from a previous 4.2~K vertical test of this cavity in a liquid helium bath \cite{porter:srf2019-mop011}, are shown in Fig. \ref{Qcompare}. 

\begin{figure}[h]
	\floatstyle{boxed}
	\begin{center}
		\includegraphics[width=1\linewidth]{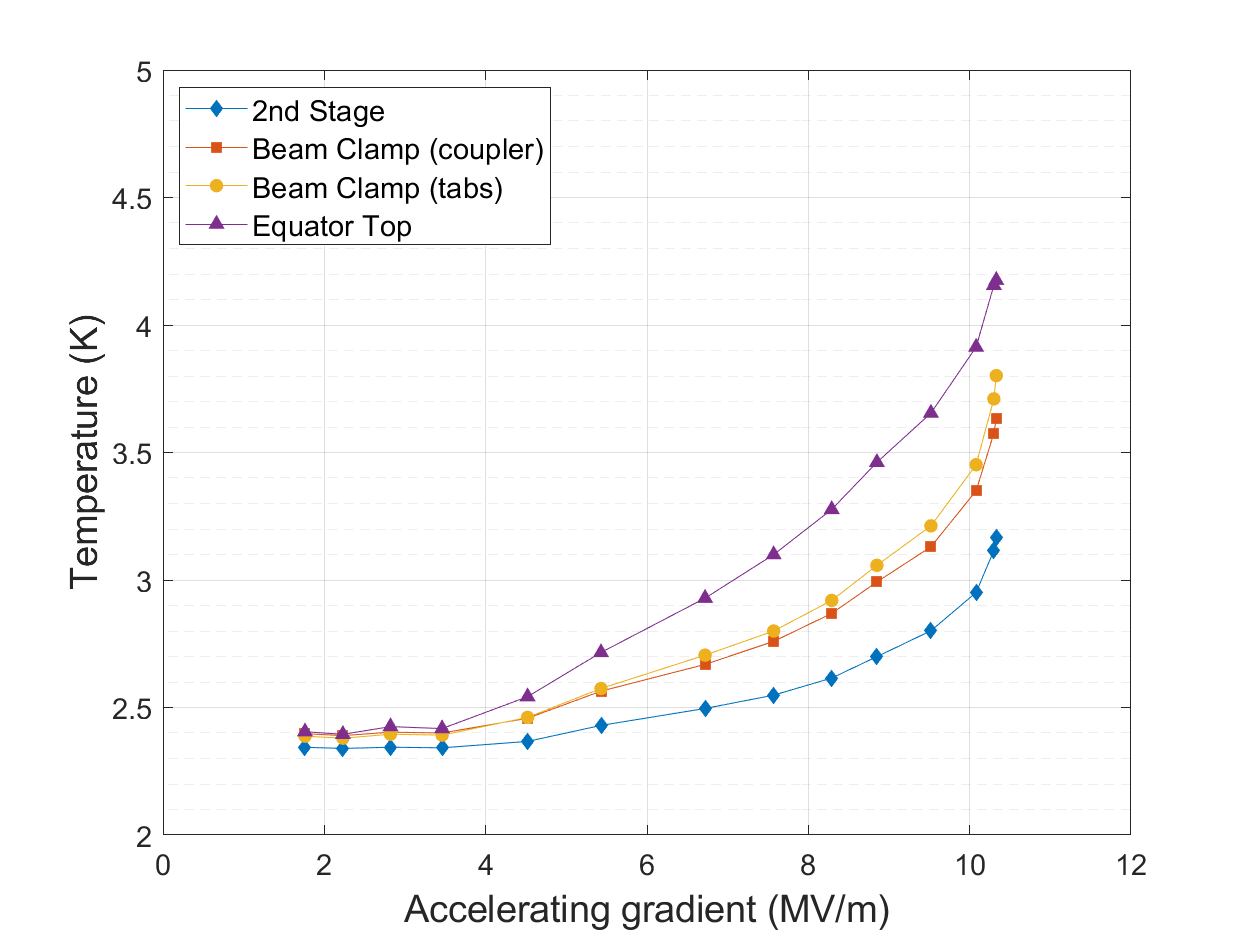}
		\caption{Temperature vs accelerating gradient from four sensors placed on the cavity assembly. All four temperatures remain close together at low fields, but at 10 MV/m there is a 1 K difference between the second stage and cavity equator temperatures. There is also a slight difference in temperature between the two beam tube clamps.}
		\label{TvE}
	\end{center}
\end{figure}

The temperature of the cavity assembly during the improved RF test after thermal cycling is shown in Fig. \ref{TvE}. The beam clamp temperatures are distinguished by the labels ``coupler'' and ``tabs,'' referring to the beam clamp on the side of the cavity closest to the forward power coupler and welded furnace tabs, respectively.  As mentioned previously, the cavity temperatures were allowed to stabilize during this test in order to represent continuous operation. Temperatures were deemed stabilized when they changed by less then 1~mK every two seconds (temperature sampling rate). The cavity saw minimal heating at low fields, with temperatures remaining under 2.5~K through 4~MV/m. At higher fields the heating became more noticeable, though the cavity remained under 4.2~K even at 10~MV/m.  A 1~K temperature difference between the cavity and second stage plate was measured at the highest fields reached.   Optimization of the thermal anchor is planned by improving thermal contacts such that this gradient can be further reduced for operation beyond 10~MV/m.

During the temperature cycling runs, the cooldown through $T_c$ was significantly faster than during the initial cooldown from room temperature (though still slow compared to fast cooldowns used in niobium SRF cavities to expel ambient magnetic fields). Tracking the cavity assembly temperatures during transition showed that the temperature gradients across the cavity cell were smaller during the 30~K cycle than the initial room temperature cooldown -- roughly 250~mK compared to~850 mK (i.e. .025~K/cm vs .085~K/cm). This indicates that the improvement in RF performance is likely due to smaller magnetic fields being created by thermoelectric currents in the bi-metal structure of the Nb$_3$Sn coated cavity, which are driven by the temperature gradients.  These additional fields are then trapped in the cavity wall when passing $T_c$, increasing losses in RF operation \cite{Hall:SRF2017-THPB042,Hall:thesis,porter:srf2019-mop011}. Moving forward, small heating elements will be added to the cavity assembly, placed on the beam tube thermal clamps to minimize temperature gradients across the cavity during cooldown.

\section{Conclusion}

We have demonstrated successful stable continuous operation of a 2.6~GHz Nb$_3$Sn cavity at 10~MV/m using conduction cooling with a commercial cryocooler. At this field, the cavity remained at 4.2~K with a quality factor of 4x10$^9$. Comparing results from tests after different temperature cycling runs indicates that temperature gradients across the cavity during the transition through T$_c$ play a significant role in affecting cavity performance. Progress for future tests will focus on reducing these thermal gradients and improving thermal conductivity from the cavity to the second stage cold head . This successful demonstration of stable operation at medium RF fields using conduction cooling marks an important first milestone in making SRF technology more  accessible to small-scale accelerators, paving the way to the use of SRF technology in a much wider range of applications.

We would like to thank the Department of Energy and National Science Foundation for providing the funding that made this research possible. U.S. DOE award DE-SC0008431 supports Nb$_3$Sn cavity coating and RF testing R\&D, including  the development of a conduction cooling test assembly, while NSF Award 1734189 supports research on compact high frequency cavities (above 2~GHz).

\bibliography{sources}

\end{document}